\documentclass{appolb}
\usepackage{epsfig}

\begin{document}
\pagestyle{plain}
\newcount\eLiNe\eLiNe=\inputlineno\advance\eLiNe by -1
\title{Charm at FAIR}
\author{L. Tol\'os$^1$\thanks{e-mail:tolos@kvi.nl}, D. Gamermann$^2$, R. Molina$^2$, E. Oset$^2$ and A. Ramos$^3$
\address{$^1$Theory Group, KVI, University of Groningen, Zernikelaan 25, 9747 AA Groningen, The Netherlands\\
$^2$Departamento de F\'{\i}sica Te\'orica and IFIC,
Centro Mixto Universidad de Valencia-CSIC,
Institutos de Investigaci\'on de Paterna, Aptdo. 22085, 46071 Valencia, Spain\\
$^3$ Departament d'Estructura i Constituents de la Mat\`eria. Universitat de Barcelona,
Diagonal 647, 08028 Barcelona, Spain}}

\maketitle

\begin{abstract}
Charmed mesons in hot and dense matter are studied within a self-consistent coupled-channel approach for the experimental conditions of density and temperature expected at the CBM experiment at FAIR/GSI. The
$D$ meson spectral function broadens with increasing density with an extended tail
towards lower energies due to $\Lambda_c(2593) N^{-1}$ and $\Sigma_c(2800) N^{-1}$ excitations. The in-medium $\bar D$ meson mass increases with density. We also discuss the consequences for the renormalized properties in nuclear matter of the charm scalar $D_{s0}(2317)$ and $D(2400)$, and the predicted hidden charm $X(3700)$ resonances at FAIR energies.
\end{abstract}

\section{Introduction}

The CBM experiment of the future FAIR project will investigate highly compressed dense matter in nuclear collisions with a beam energy range between 10 and 40 GeV/u \cite{fair}. One of the goals is to extend the SIS/GSI program for the in-medium modification of hadrons and to provide first insight into charm-nucleus interaction. The study of the properties of elementary particles in nuclei helps to learn about not only the excitation mechanisms in the nucleus but also the properties of these particles. Thus, the possible modifications of the properties of open and hidden charmed mesons in a hot and dense environment are  matter of recent analysis.

\begin{figure}[htb]
\begin{center}
\includegraphics[height=5.5 cm, width=6 cm]{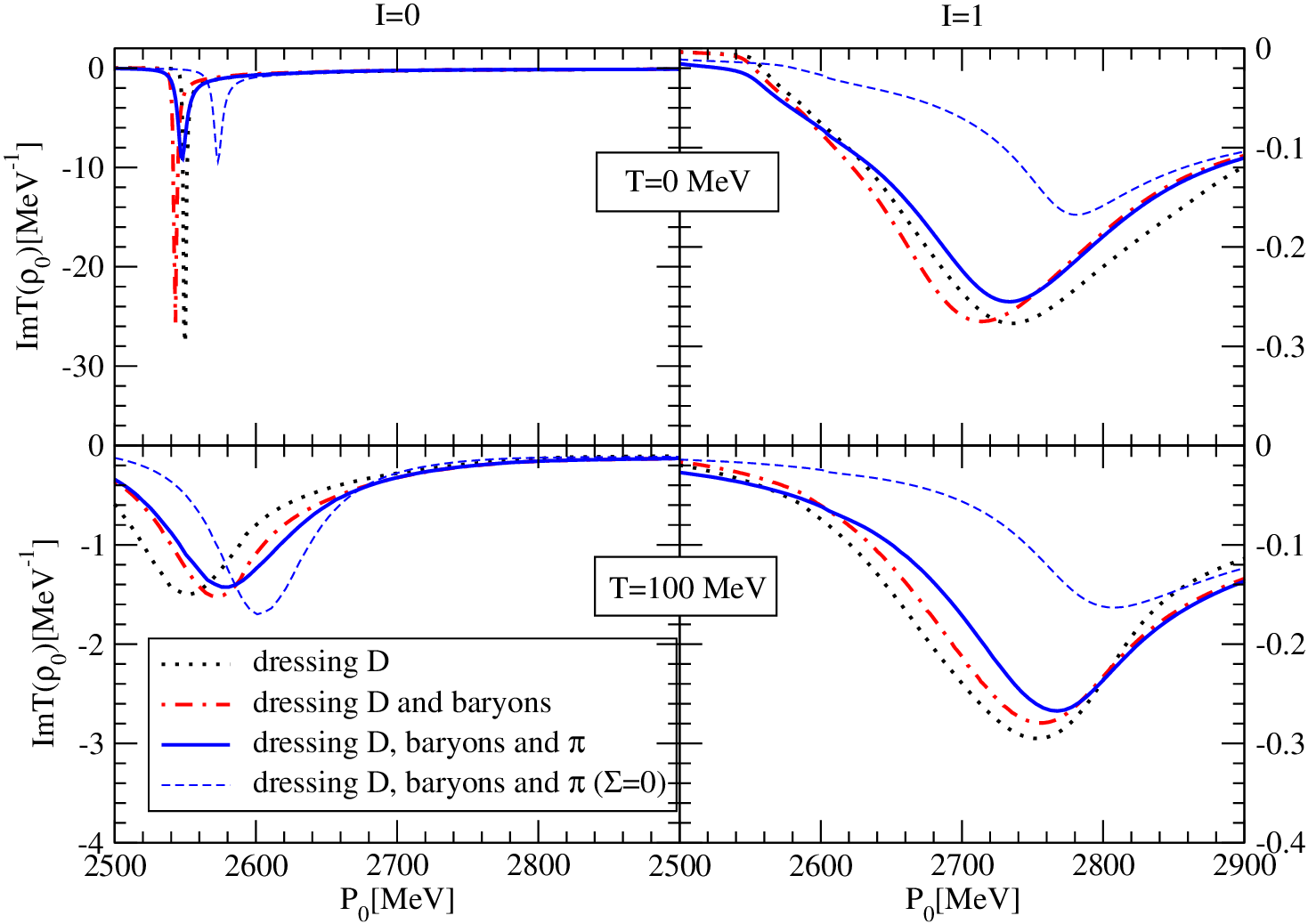}
\hfill
\includegraphics[height=5.5 cm, width=6 cm]{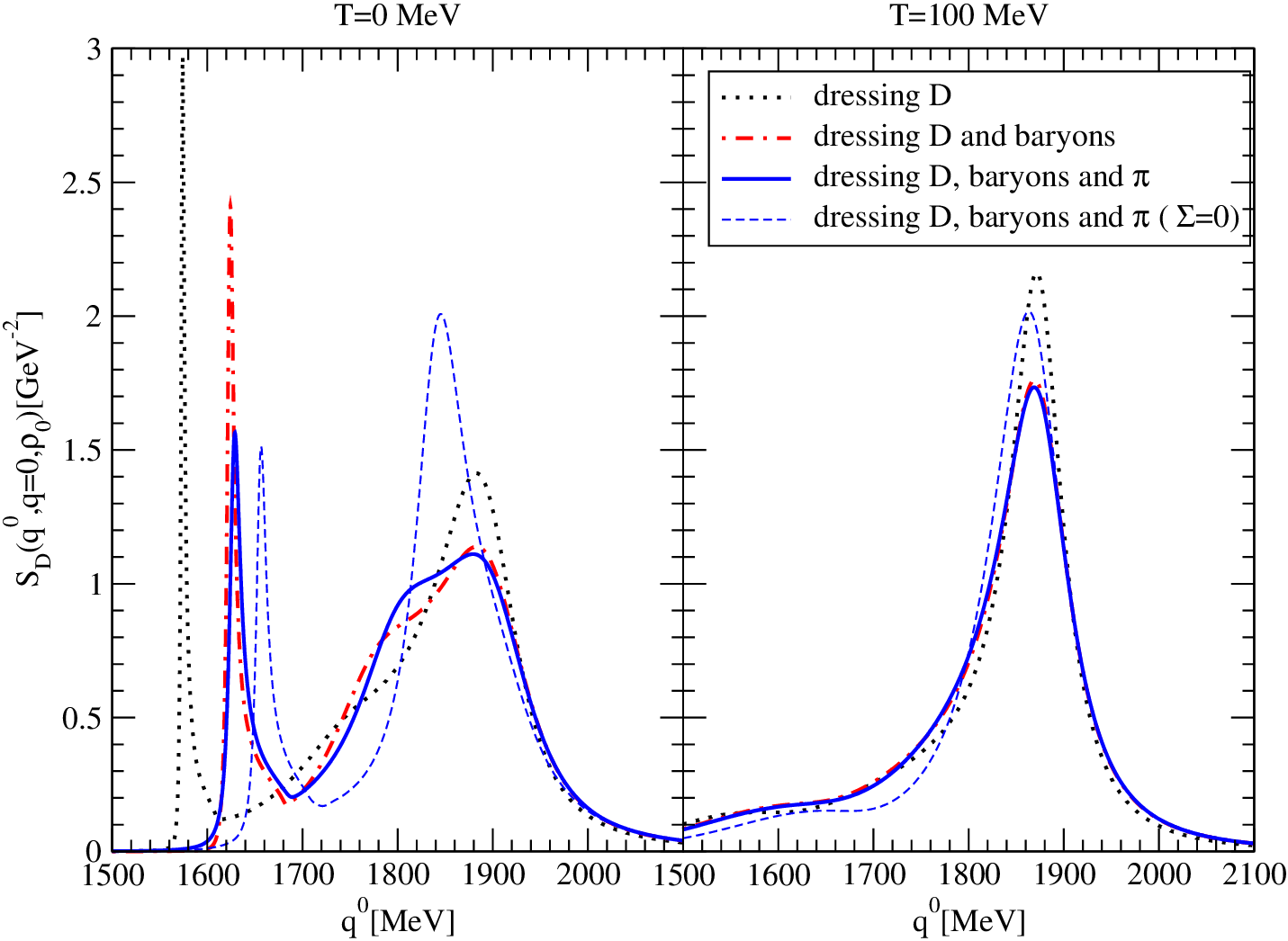}
\caption{$\tilde{\Lambda}_c$ and $\tilde{\Sigma}_c$ resonances, and the $D$ meson spectral function} \label{fig:DN}
\end{center}
\end{figure}

The in-medium modification of the open charm mesons ($D$ and $\bar D)$ may explain the $J/\Psi$ suppression \cite{MAT86} in an hadronic environment,  based on the mass reduction of $D (\bar D)$ in the nuclear medium. However, a coupled-channel meson-baryon
scattering in nuclear medium is needed due to the strong coupling among the
$DN$ and other meson-baryon channels \cite{TOL04,TOL06,LUT06,MIZ06, TOL07}. Moreover, changes in the properties of open charm will affect the renormalization of charm and hidden charm scalar mesons in nuclear matter, providing some information about their nature, whether they are $q\bar{q}$ states, molecules, mixtures of $q\bar{q}$ with meson-meson components, or dynamically generated resonances resulting from the interaction of two pseudoscalars.

In the present article, we pursue a coupled-channel study on the spectral properties of $D$ and $\bar D$ mesons in nuclear matter at finite temperature. We then analyze the effect of the self-energy of $D$ mesons on dynamically-generated charm and hidden charm scalar resonances, such as $D_{s0}(2317)$ and $D(2400)$, and the predicted hidden charm $X(3700)$.

\section{Charm mesons in a hot nuclear environment}
\label{section1}

\begin{figure}[htb]
\begin{center}
\includegraphics[height=4.5 cm, width=5 cm]{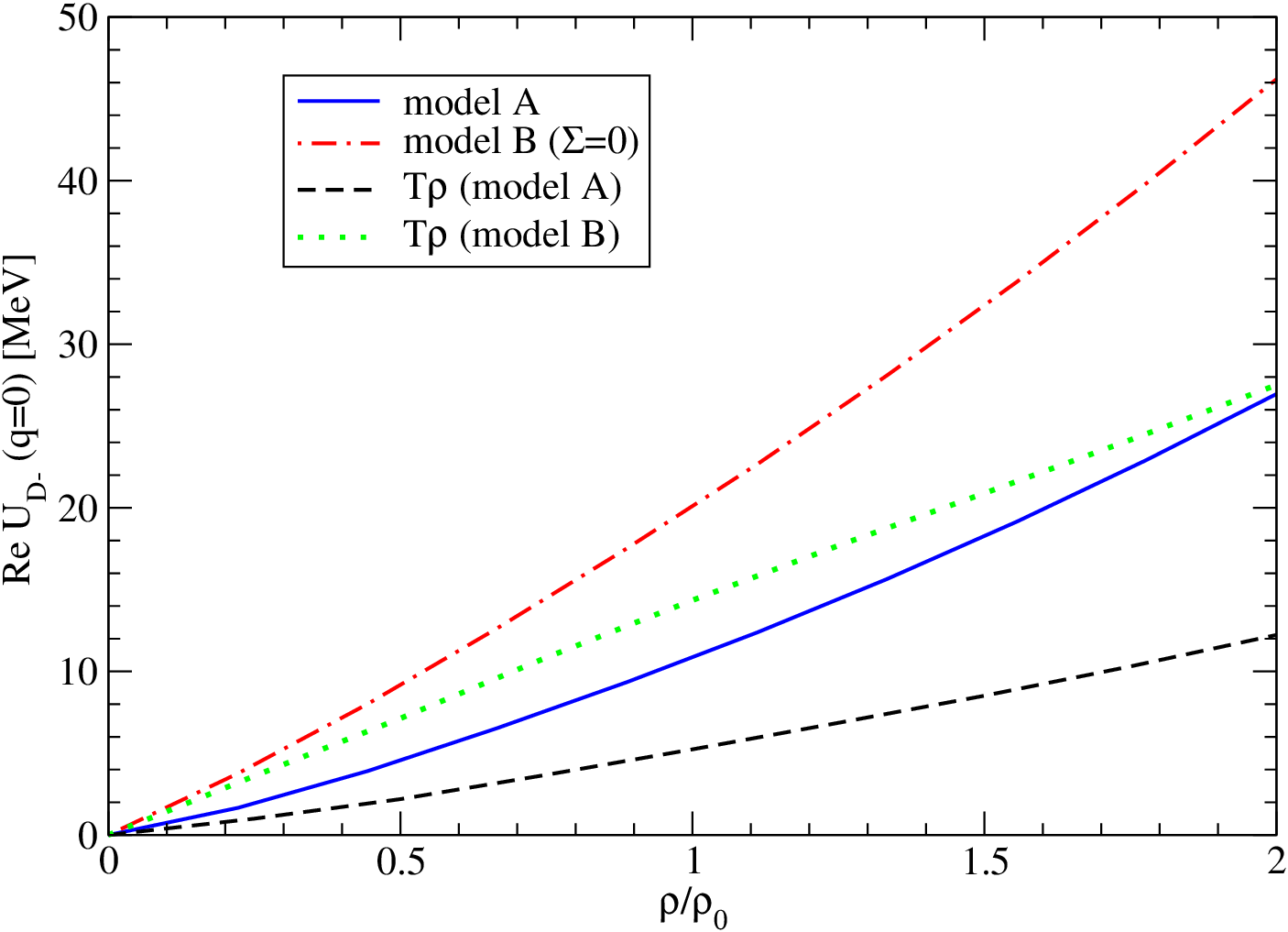}
\hfill
\includegraphics[height=5.5 cm, width=7 cm]{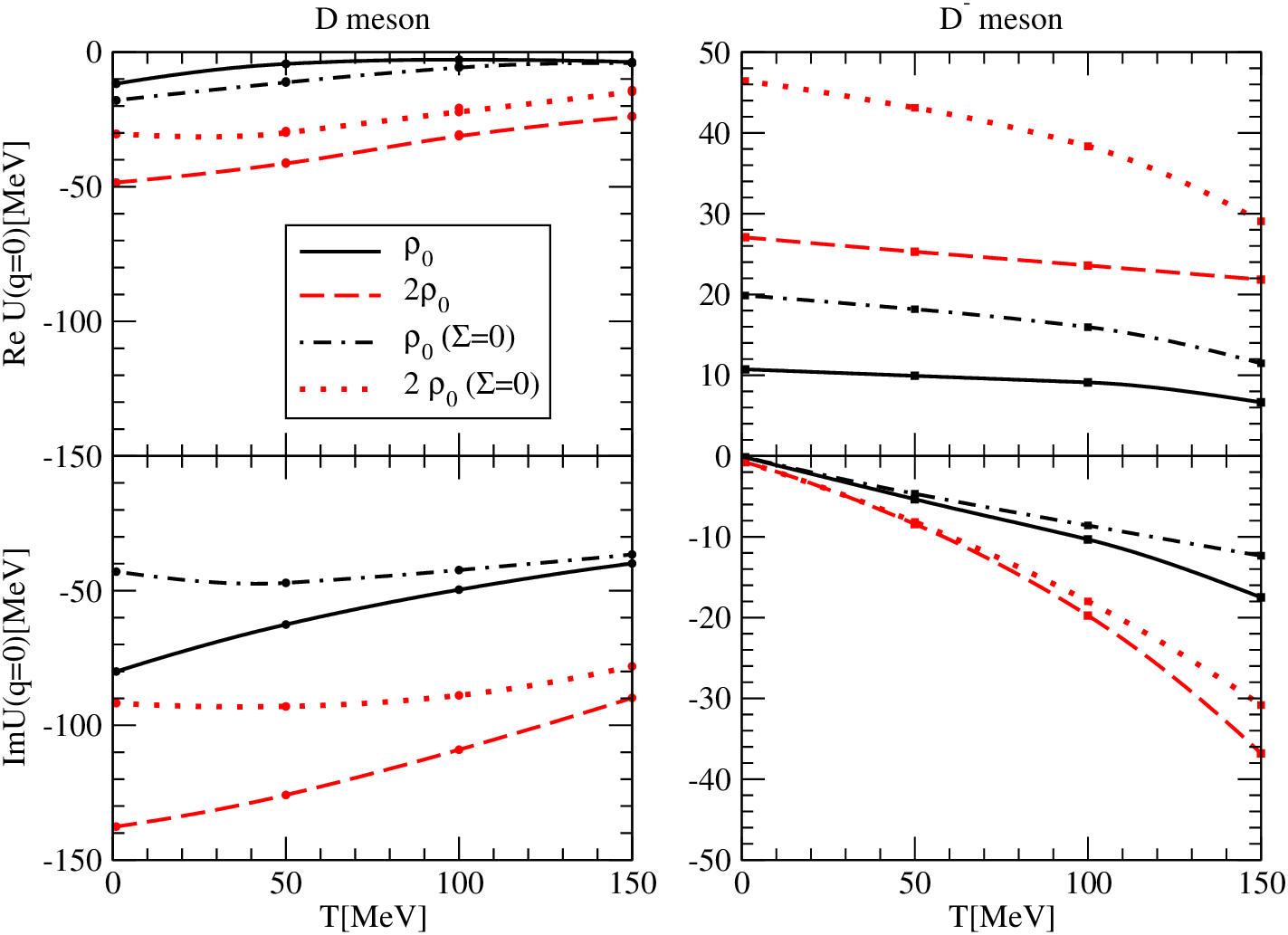}
\caption{$\bar D$ mass shift as well as $T\rho$, and the $D$ and $\bar D$ potentials.} \label{fig:DbarN}
\end{center}
\end{figure}

We study the spectral properties of $D$ and $\bar D$ mesons in nuclear matter at finite temperature by extending the result of Ref.~\cite{MIZ06}. The $D$ and $\bar D$ self-energies at finite temperature are obtained from a self-consistent coupled-channel calculation taking, as bare interaction, a type of broken SU(4) $s$-wave Tomozawa-Weinberg (TW) interaction supplemented by an attractive isoscalar-scalar term ($\Sigma_{DN}$). The transition matrix $T$ is solved using a cutoff regularization, which is fixed by reproducing the position and the width of the $I=0$ $\tilde{\Lambda}_c(=\Lambda_c(2593))$ resonance while a new resonance in $I=1$ channel $\tilde{\Sigma}_c(=\Sigma_c(2800))$ is then generated \cite{MIZ06}.

The in-medium solution at finite temperature incorporates Pauli blocking effects, baryon mean-field bindings via a temperature-dependent $\sigma -\omega$ model, and $\pi$ and open-charm meson self-energies in the intermediate propagators (see \cite{TOL07}). The self-energy and, hence, spectral function are obtained self-consistently summing $T_{DN}$ over the nucleon Fermi distribution.

The $I=0$
$\tilde{\Lambda}_c$ and $I=1$ $\tilde{\Sigma}_c$ resonances in hot dense matter are shown in the l.h.s. of Fig.~\ref{fig:DN} for three different self-consistent calculations:  i) including only the
self-consistent dressing of the $D$ meson, ii) adding the mean-field binding of baryons (MFB) and iii) 
including MFB and the pion self-energy (PD). The thick lines correspond to model A (viz.
$\Sigma_{DN} \neq 0$) while the thin-dashed lines refer to Case (iii) within model B ($\Sigma_{DN}=0$). Medium effects at $T=0$ lower the
position of the $\tilde{\Lambda}_c$ and $\tilde{\Sigma}_c$ with respect
to their free  values. Their width values, which increase due to $\tilde
Y_c (=\tilde \Lambda_c, \tilde \Sigma_c) N \rightarrow \pi N \Lambda_c, \pi N \Sigma_c$  processes, differ according to the phase space available. The PD induces a small effect in the resonances because of charm-exchange channels being suppressed, while models A and B are qualitatively similar. Finite temperature results in the reduction of the Pauli blocking due to the smearing of the Fermi surface  with temperature. Both resonances move up in energy closer to their free position while they are smoothed out, as in \cite{TOL06}.

In the r.h.s of Fig.~\ref{fig:DN} we display the $D$ meson spectral function for (i) to (iii) (thick lines) for model A and  case (iii) for model B (thin line). At $T=0$ the spectral function presents two peaks: $\tilde \Lambda_c N^{-1}$ excitation at a lower energy whereas the second one
at higher energy is the quasi(D)-particle  peak  mixed with  the $\tilde \Sigma_c N^{-1}$ state. Once MFB is included, the lower peak built up by the $\tilde \Lambda_c N^{-1}$ mode goes up by about $50$ MeV relative to (i) since the meson requires to carry more energy to compensate for the attraction felt by the nucleon.  The same characteristic feature is seen for the $\tilde \Sigma_c N^{-1}$ configuration that mixes with the quasiparticle peak. The PD does not alter much the position of $\tilde \Lambda_c
N^{-1}$ excitation or the quasiparticle peak. For model B ((iii) only), the absence of the $\Sigma_{DN}$ term moves the $\tilde \Lambda_c N^{-1}$ excitation closer to the  quasiparticle peak, while the latter fully mixes with the $\tilde \Sigma_c N^{-1}$ excitation. At finite temperature those structures dilute with increasing temperature while the quasiparticle peak gets closer to its free value becoming narrower, because the self-energy
receives contributions from higher momentum $DN$ pairs where the interaction is weaker.

In the $\bar D N$ sector, the scattering lengths for model A (B) are $a^{I=0}=0.61 \ (0)$ fm and   $a^{I=1}=-0.26 \ (-0.29)$ fm.  While our repulsive $I=1$ is 
in good agreement with \cite{LUT06}, the finite
value for the $I=0$ scattering length found in this latter reference is in
contrast to the zero value found here for model B due to the vanishing
$I=0$ coupling coefficient of the corresponding pure TW  $\bar
DN$ interaction. Our results are, however, consistent
with Ref.~\cite{HAI07}. For model A, we obtain a non-zero value of the $I=0$ scattering length,
due to the magnitude of the $\Sigma_{DN}$ term. As seen in the l.h.s of Fig.~\ref{fig:DbarN}, the $\bar D$ mass shift in cold nuclear matter is repulsive and, in spite of
the absence of resonances close
to threshold, the low-density approximation or $T \rho$ breaks down at normal nuclear matter density $\rho_0=0.17  \ {\rm fm}^{-3}$.

Finally, in the r.h.s of  Fig.~\ref{fig:DbarN} we compare the $D$ and
$\bar D$ optical potentials. For model A (B) 
at $T=0$, we obtain an attractive potential of $-12$ ($-18$) MeV for the
$D$ meson, similar to \cite{TOL06}, while the repulsion for $\bar D$ is $11$ ($20$) MeV. The temperature dependence of the repulsive real part of the ${\bar D}$ optical
potential is very weak, while the imaginary part increases steadily due to the
increase of collisional width. The picture is somewhat different for the $D$
meson due to the overlap of the quasiparticle peak with the $\tilde{\Sigma}_c N^{-1}$ mode. The $\tilde{\Sigma}_c N^{-1}$ mode also alters the effect of the $\Sigma_{DN}$ term on the potential.


\section{Charm and hidden charm scalar resonances}

\begin{figure}[htb]
\begin{center}
\includegraphics[height=5.5 cm, width=4 cm]{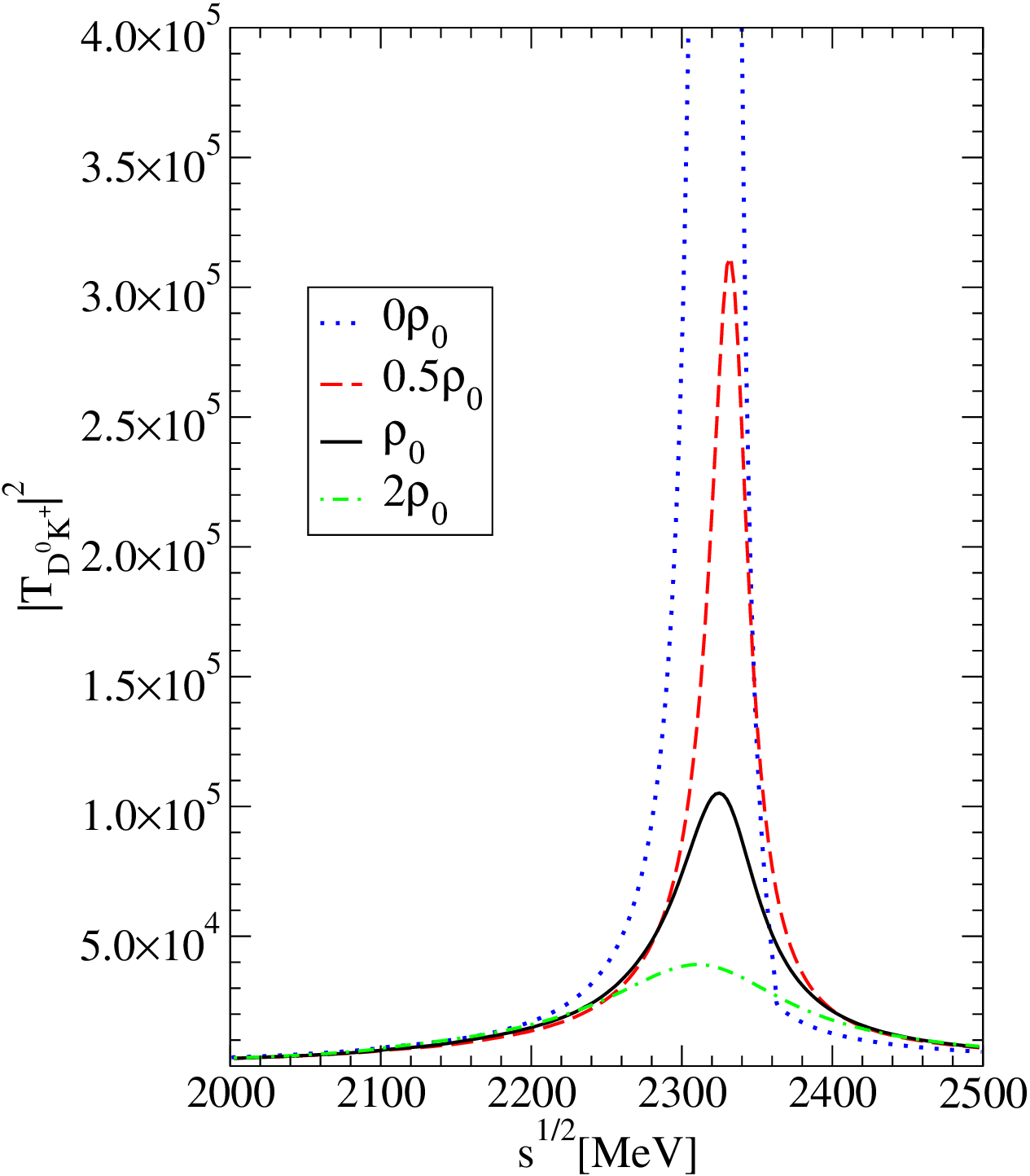}
\hfill
\includegraphics[height=5.5 cm, width=4 cm]{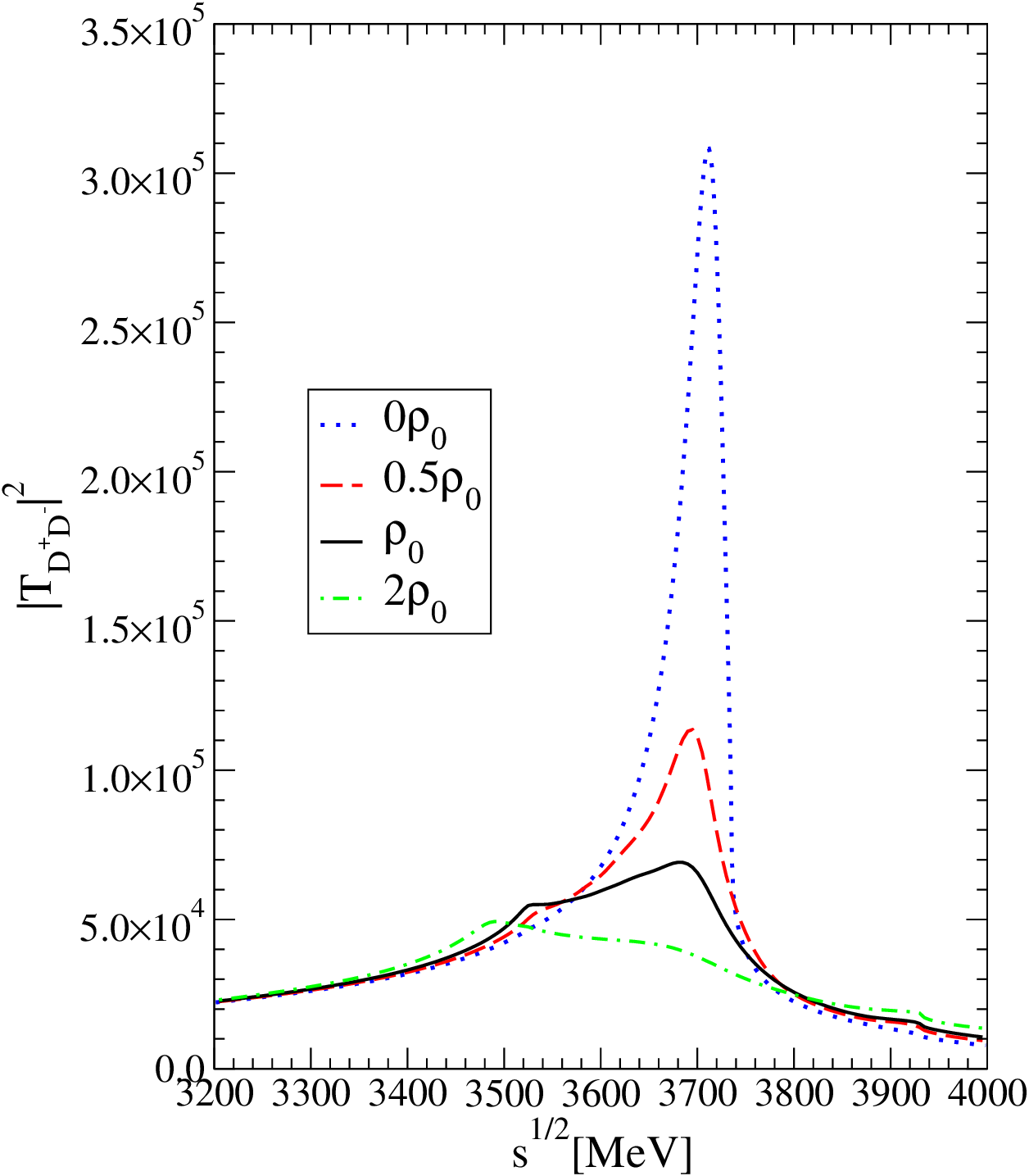}
\hfill
\includegraphics[height=5.5 cm, width=4 cm]{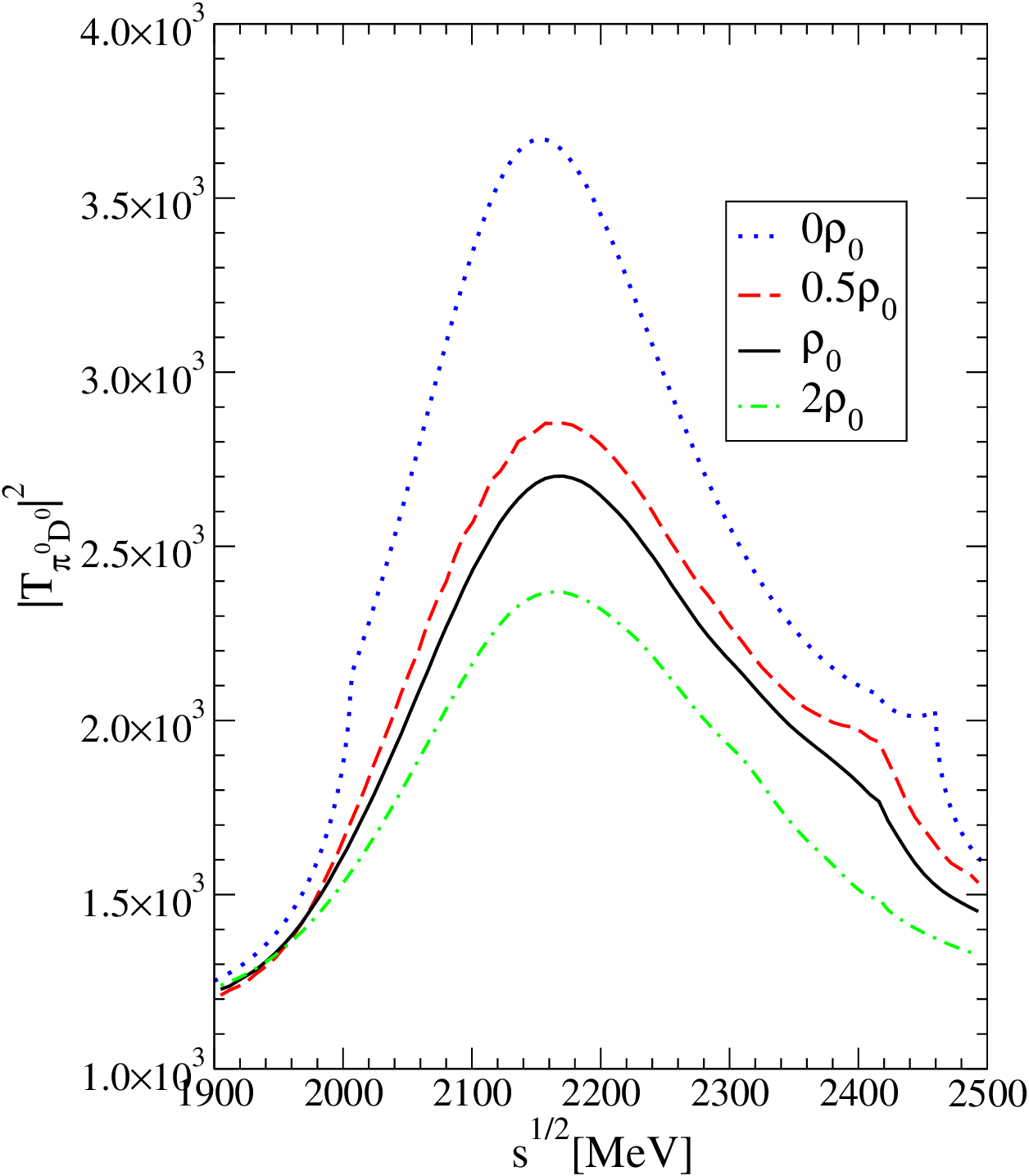}
\caption{$D_{s0}(2317)$ (left), $X(3700)$ (middle) and $D_0(2400)$ (right) resonances.} \label{fig2}
\end{center}
\end{figure}


Establishing the nature of a resonance, whether it has the usual $q \bar q$/$qqq$ structure or is better described as being dynamically generated, is an active matter of research, in particular, for scalar resonances. Via their renormalized properties in nuclear matter we can not only learn about the excitation mechanisms in the nucleus but also the properties of these particles. 

We study the charmed resonances $D_{s0}(2317)$ and $D(2400)$ \cite{Kolomeitsev:2003ac,Guo:2006fu,Gamermann:2006nm} together with a hidden charm scalar meson, $X(3700)$, predicted in \cite{Gamermann:2006nm}, which might have been observed by the Belle collaboration \cite{Abe:2007sy} via the reanalysis of \cite{Gamermann:2007mu}. Those resonances are generated dynamically solving the coupled-channel Bethe-Salpeter equation for two pseudoscalars \cite{Molina:2008nh}. The kernel is derived from a $SU(4)$ extension of the $SU(3)$ chiral Lagrangian used to generate scalar resonances in the light sector. The $SU(4)$ symmetry is, however, strongly 
 broken, mostly due to the explicit consideration of the masses of the vector 
 mesons exchanged between pseudoscalars \cite{Gamermann:2006nm}. 

The analysis of the transition amplitude close to each resonance for the different coupled channels give us information about the coupling of the resonance to a particular channel. The $D_{s0}(2317)$ mainly couples to $DK$ system, while the $D_0(2400)$ to $D\pi$ and, secondly, to $D_s \bar K$.  And the  hidden charm state $X(3700)$ couples most strongly to 
$D\bar{D}$. Therefore, any change in the $D$ meson properties in nuclear matter will have an important effect on these  resonances. Those modifications are given by the $D$ meson self-energy in nuclear matter, as discussed in Sec.~\ref{section1}, but supplemented by the $p$-wave self-energy through the corresponding $Y_cN^{-1}$ excitations \cite{Molina:2008nh}.

 In Fig.~\ref{fig2}, the resonances $D_{s0}(2317)$ and $D_0(2400)$  and $X(3700)$ are shown by displaying the squared transition amplitude for the corresponding dominant channel at different densities. In the case of the $D_{s0}(2317)$ and 
$X(3700)$ resonances, which have a zero and small width, respectively,
the medium effects lead to widths of the order of 100 and 200
MeV at normal nuclear matter density, correspondingly. The origin can be traced back to the opening of new many-body decay channels, as the $D$ meson gets absorbed in the nuclear medium via $DN$ and $DNN$ inelastic reactions.  For the $D_0(2400)$, we observe an 
 extra widening from the already large width of the resonance in free space. However,
  the large original width makes the medium effects comparatively much 
  weaker than for the other two resonances \cite{Molina:2008nh}. In our model, we do not extract any clear conclusion for the mass shift. We suggest to look at transparency ratios to investigate those in-medium widths. This magnitude, which gives the survival probability in production reactions in  nuclei, is very sensitive to the absorption rate of any resonance inside nuclei, i.e., to its in-medium width.

\section{Summary and Outlook}

We have studied the properties of $D$ and $\bar D$ mesons within a self-consistent coupled-channel approach for the experimental conditions expected at the CBM experiment at FAIR. The in-medium $\bar D$ mass increases by about $10-20$ MeV whereas the $D$ spectral function extends to lower "mass" due to the thermally spread $\tilde Y_c N^{-1}$. However, it is unlikely to explain $J/\Psi$ suppression via the $D \bar D$ decay in hot dense matter. A more plausible hadronic contribution for the $J/\Psi$ suppression is the reduction of its supply from the excited charmonia, $\chi_{c\ell}(1P)$ or $\Psi'$, which may find in the medium other competitive decay channels. We have also evaluated the renormalized properties in nuclear matter of the charm scalar $D_{s0}(2317)$ and $D(2400)$, and the predicted hidden charm $X(3700)$ resonances. Those resonances develop an important width in this dense environment. We conclude that the experimental analysis of those properties is a valuable test of the dynamics of the $D$ meson interaction with nucleons and
nuclei, and the nature of those charm and hidden charm scalar resonances. Altogether, those results should stimulate experimental work in hadron facilities, in particular at FAIR \cite{fair}, where the charm degree of freedom will be throughly investigated.

\end{document}